\documentclass[conference]{IEEEtran}
\ifCLASSINFOpdf
\else
\fi

\usepackage{array}
\usepackage{amsmath}
\usepackage{amssymb}
\usepackage{setspace}
\usepackage{graphicx}
\usepackage{epstopdf}
\usepackage[tight,footnotesize]{subfigure}
\usepackage{algorithm}
\usepackage{algorithmicx}
\usepackage{algpseudocode}
\usepackage{subfigure}
\usepackage{float}
\usepackage{cite}

\begin{document}

\title{An Efficient Privacy-Preserving Incentive Scheme without TTP in Participatory Sensing Network}

\author{\IEEEauthorblockN{Jingwei Liu\IEEEauthorrefmark{1},
Xiaolu Li\IEEEauthorrefmark{1},
Rong Sun\IEEEauthorrefmark{1},
Xiaojiang Du\IEEEauthorrefmark{2} and
Paul Ratazzi\IEEEauthorrefmark{3}}
\IEEEauthorblockA{\IEEEauthorrefmark{1}State Key Lab of ISN, Xidian University, Xi'an, 710071, China.\\ Email: jwliu@mail.xidian.edu.cn, lixiaolu0318@163.com, rsun@mail.xidian.edu.cn}
\IEEEauthorblockA{\IEEEauthorrefmark{2}Department of Computer and Information Sciences, Temple University, Philadelphia, PA 19122, USA.\\
Email: dxj@ieee.org}
\IEEEauthorblockA{\IEEEauthorrefmark{3}Air Force Research Laboratory, Rome, NY, USA.\\
Email: edward.ratazzi@us.af.mil}
}
\maketitle

\begin{abstract}
Along with the development of wireless communication technology, a mass of mobile devices are gaining stronger
sensing capability,  which brings a novel paradigm to light: participatory sensing networks (PSNs). PSNs can greatly reduce the cost of wireless sensor networks, and hence are becoming an efficient way to obtain abundant sensing data from surrounding environment. Therefore, PSNs would lead to significant improvement in various fields, including cognitive communication. However, the large-scale deployment of participatory sensing applications is hindered by the lack of incentive mechanism, security and privacy concerns. It is still an ongoing issue to address all three aspects simultaneously in PSNs. In this paper, we construct an efficient privacy-preserving incentive scheme without trusted third party (TTP) for PSNs to motivate user-participation. This scheme allows each participant to earn credits by contributing data privately. Using blind and partially blind signatures, the proposed scheme is proved to be secure for privacy and incentive. Additionally, the performance evaluation in terms of computation and storage indicates that the proposed scheme has higher efficiency.
\end{abstract}
\IEEEpeerreviewmaketitle

\section{Introduction}

In recent years, mobile phones have made great progress in processing power, storage capacities, embedded sensors (e.g., accelerometer, GPS, microphone), and communication capabilities. The rapid penetration of mobile phones not only changed the traditional internet service model, but also increased the diversity of applications. One of the applications is based on the data collected by the sensors in the powerful mobile phones, the emerging application---participatory sensing networks (PSNs) \cite{burke2006participatory}. PSNs are greatly helpful to the improvement of cognitive communication that combines perception, information processing, artificial intelligence and machine learning together.


\begin{figure}[tb]
\begin{center}
\includegraphics[width=8.5cm]{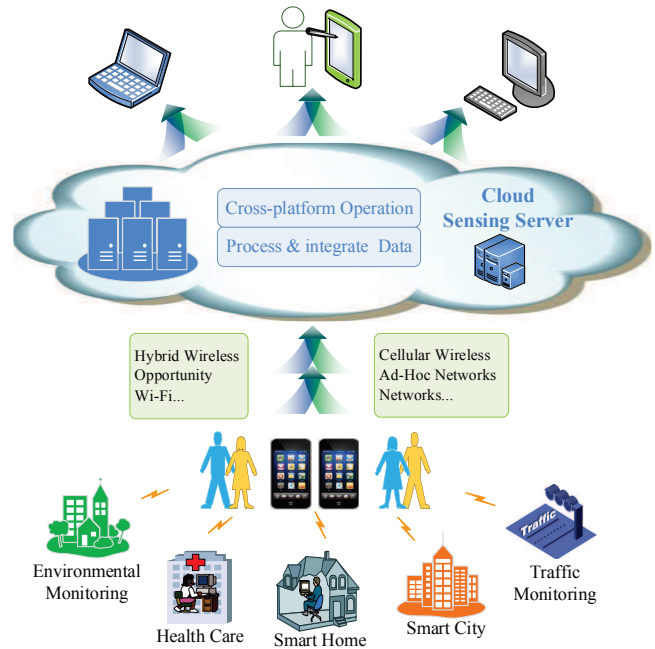}\\
\caption{A Basic framework of PSNs}\label{application}
\end{center}
\end{figure}

A PSN system is essentially a wireless sensor network (WSN) formed by ubiquitous sensors. However, compared with traditional WSNs, sensors are no need to pre-distributed in PSNs, which reduces setup cost. Moreover, people-centric PSNs provide better spatial and temporal coverage. A basic framework of PSNs is shown in Fig. \ref{application}. Participants collect sensing data through the sensors embedded in smart terminals and upload these information to a cloud server. The server integrates and analyzes all the sensing data, then shares the results with the corresponding customers. Therefore, participants are not only the data providers in PSNs, but also the consumers and ultimate beneficiaries.

Since the concept of PSN was initially presented in 2006, it has been widely applied to environmental monitoring \cite{mun2009peir}, traffic route navigation \cite{thiagarajan2009vtrack}, health care \cite{hicks2010andwellness}, etc. In addition, PSNs have made a greater contribution in participatory cognitive radio networks \cite{nadendla2012auction}. However, in various of application scenarios, PSNs have to face to many security challenges \cite{xiang2011low, du2005maintaining, xu2017secure, xiao2007survey, du2008secure, liu2016smart, du2009transactions, zhang2018connecting, du2007effective, yu2016feasible, du2006secure, xiao2007internet, hu2005optimized, du2008security}. Although there are some methods \cite{yu2012discriminating, zhang2013network, manandhar2014detection, yu2011traceback}, the critical challenge that the contradiction between privacy preservation and incentive mechanism, which hinders the large-scale deployment of mobile sensing applications. Without reasonable and secure reward, participants may not be willing to spend time, effort or money on any sensing task. Therefore, an appropriate incentive mechanism is necessary to stimulate participants' enthusiasm and persistence. However, the identities and other sensitive information of participants may be abused by the cloud server in the incentive scheme. Therefore, in this paper, we propose an efficient privacy-preserving incentive scheme to meet requirement of security and privacy preservation in PSNs.

Although there have been plenty of research efforts on privacy preservation in PSNs \cite{pingley2009cap, de2011short, cornelius2008anonysense}, most of them do not consider incentive mechanisms for participants. Cristofaro et al. \cite{de2011short} proposed a secure framework of participant-sensing instead of the detailed algorithm. Kapadia et al. \cite{cornelius2008anonysense} proposed a scheme to gather sensing data anonymously, named AnonySense. Nevertheless, in these schemes, it is still an open issue that there is few appropriate incentives involved in the sensing task to attract more users' participation.

In addition, many studies of incentive mechanisms in PSNs have appeared gradually. In \cite{lee2010sell}, Lee et al. proposed a reverse auction scheme based on dynamic price, named RADP. Yang et al. \cite{yang2012crowdsourcing} presented two incentive models based on the games and auctions from the perspective of data requestors and participants separately. Nonetheless, there are no privacy-preserving measures in these schemes, so the users' privacy is likely to be leaked. It may cause unnecessary troubles.

Until now, the joint-design on the above two issues still has not attracted sufficient research attention. They urgently need to be considered for adapting the extensive applications of PSNs. For instance, one of the most effective solutions, the pseudonym, is used to conceal a participant's real identity in PSNs. In this scheme, each participant generates his/her tokens and commitments in cooperation with the server using blind and partially blind signature to protect privacy against the attacks by any third party. In this paper, we propose an efficient privacy-preserving incentive scheme without trusted third party (TTP) through a credit-based approach, which allows each user to earn credits by contributing sensing data without leaking his/her privacy.

The rest of this paper is organized as follows. In section II, we briefly introduce some preliminaries including the system model, threats models, cryptographic primitives, etc. In section III, we describe the proposed scheme in detail and analyze its security properties. In section IV, the performance is evaluated. Finally, we conclude the paper in section V.
\section{Preliminaries}
\subsection{System Model}

\begin{figure}[tb]
 \centering
  \includegraphics[width=8cm]{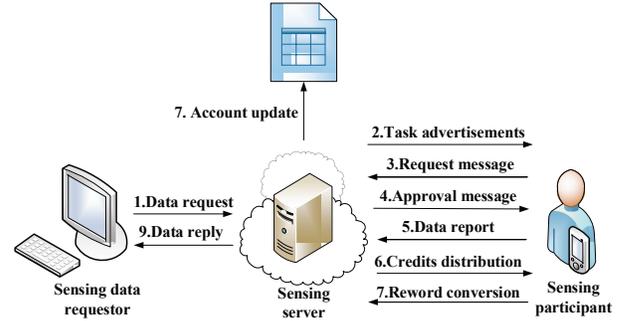}\\
 \caption{System model}
 \label{model}
\end{figure}

Recently, some models have been proposed for data collection or processing \cite{xu2016analytical, baek2015secure}. In this paper, we propose a people-centric model for the privacy-preserving incentive scheme, as shown in Fig. \ref{model}. We define three different entities that are identified as follows:

\begin{itemize}
\item Sensing Data Requestor (SDR): The SDRs send queries to the sensing server for the desired statistics and context data. As customers of PSN services, they need to indicate which types of data they are interested in obtaining.\par
\item Sensing Participant (SP): The SPs are responsible for collecting the relevant data with sensors and uploading them to the sensing server via 3G/4G or Wi-Fi.\par
\item Sensing Server (SS): The SS manages the PSN services to facilitate effective sharing of data between SPs and SDRs. The SS collects sensing reports from the SPs and answers the SDRs based on analysis results on the collected data.\par
\end{itemize}

The basic workflow is described as follows. When an SDR requires some data, it needs to send a query to the SS. The SS transforms the query into one or more tasks and publishes them to a task queue. If the SP decides to take part in a certain task, he/she will collect sensing data in accordance with the task requirements and summarize a report with a random pseudonym. Then, he/she submits the report to the SS using a new pseudonym. Accordingly, the SS pays a certain number of credits to the corresponding SP. After obtaining enough data reports, the SS integrates and analyzes all reports, then sends feedback to the SDR.

\subsection{Threat Models}

1) Threats to incentive

With respect to incentive, we assume that the SS is honest. On receiving valid virtual credits from SPs, the SS will not repudiate to pay. Otherwise, it would affect SPs' enthusiasm to participate in the sensing task in the future. In this case, we assume there exist two types of threats to the incentive mechanisms: firstly, a dishonest SP may upload the same data report repeatedly or reuse expired credit tokens in an attempt to obtain more credits than allowed; secondly, a malicious SP may compromise the other SPs for their credit tokens, or forge the tokens to earn more credits.

2) Threats to privacy

We assume that the SS may be curious about which tasks have been accepted or which reports have been submitted by SPs. Thus, there are also two types of threats to privacy of the SPs in this case. On one hand, the SS tries to link the identities of participants to some reports that may contain sensitive information. On the other hand, a malicious server may attempt to infer a certain participant' real identity through multiple tasks requested by him/her.

\subsection{Cryptographic Primitives}
\subsubsection{Pseudonym}
Due to the pseudonym, no entity can link a submitted report to the actual participant. To the best of our knowledge, the pseudonym is an effective solution to protect SPs' privacy. It is difficult for an adversary to establish a relationship between a participant and the data report from a randomized ID instead of the actual ID.
\subsubsection{Blind Signature}
Blind signature was first introduced by D. Chaum in 1983, which can effectively protect the content of the messages that need to be signed. Due to properties of blindness, untraceability and unforgeability, blind signature is often used in electronic cash protocols to protect privacy.
\subsubsection{Partially blind signature}
In partially blind signature, there is a part of common information should be pre-agreed upon (e.g., index of task). Similarly, the signer is not allowed to know the other part of information. Thus, he/she cannot link the signature to the communication session from which the signature is obtained.
\subsection{Assumptions}
In this paper, we put forward following three assumptions:

\begin{itemize}
\item Firstly, we assume that the SS and each SP have several pairs of public/private keys issued by a certified authority to authenticate each other.
\item Secondly, we suppose that each task can only be requested once by each participant.
\item Thirdly, we assume that users' mobile phones should be kept securely.
\end{itemize}
\section{An Efficient Privacy-Preserving Incentive Scheme without TTP in PSNs}

 To solve the contradiction between privacy preservation and incentive mechanism, we propose an efficient privacy-preserving incentive scheme without TTP for PSNs. In the scheme, we combine three types of techniques: pseudonyms, blind signature and partially blind signature together to protect the privacy of SPs.

\subsection{Design Objectives}
The proposed scheme should provide a reasonable incentive mechanism to maintain users' enthusiasm for their participation, and ensure that attackers cannot get users' sensitive information. In terms of incentive, SPs can obtain corresponding credits by completing the sensing tasks. In general, an SP can earn at most $c_{max}$ credits from the task published by the SS. With respect to privacy-preserving, when the SP requests multiple tasks, the SS is unable to link these tasks to it. Also, the SS cannot link the uploaded reports to it.

\subsection{Privacy-Preserving Incentive Scheme in PSNs}
In PSNs, the SS may publish many tasks in batch in a certain time slot, namely task windows. In chronological order, various tasks are distributed in these windows. Without loss of generality, we consider the first task window. The tasks in this window are numbered $1,2,...,M$. The SS needs to maintain a list of tokens for each task.
\begin{table}
  \centering
  \caption{Notations} \label{notations}
  \footnotesize
  \tabcolsep 0.05in
  \setlength{\extrarowheight}{0.1cm}
 \begin{tabular}{l|l}
    \hline
     \raisebox{0.0cm}{Notations}   & \raisebox{0.0cm}{Description}  \\
    \hline
     \raisebox{0.0cm}{$e,d$}  &  \raisebox{0.0cm}{The public/private keys for blind RSA signature}    \\
     \raisebox{0.0cm}{$K_1,K_2$}  &  \raisebox{0.0cm}{The SS's private keys for partially blind signature}   \\
     \raisebox{0.0cm}{$r_1,r_2,r_3$}  &  \raisebox{0.0cm}{The secure numbers of SP}    \\
     \raisebox{0.0cm}{$c_{i}\in[c_{min},c_{max}]$}  &  \raisebox{0.0cm}{The number of credits paid to SP for a task}   \\
     \raisebox{0.0cm}{$\gamma_i, \delta_{ij}, \varepsilon_{ij}$}  &  \raisebox{0.0cm}{Request token, report token, credit token} \\
     \raisebox{0.0cm}{$H$}  &  \raisebox{0.0cm}{A cryptographic hash function}  \\
     \raisebox{0.0cm}{$T_i$}  &  \raisebox{0.0cm}{A timestamp}   \\
     \raisebox{0.0cm}{$RID,PID$}  &  \raisebox{0.0cm}{The real identity and pseudonym of an SP}  \\
    \hline
 \end{tabular}
\end{table}

In the proposed scheme, we use blind and partially blind signature in \cite{zhang2003efficient} to protect SPs' sensitive information. The primary notations used in this paper are given in TABLE \ref{notations}. The SS generates the private/public key pair $d$ and $e$ for RSA signature and blind RSA signature, two private keys $K_1, K_2$ for partially blind signature. Each SP randomly selects three secure numbers: $r_1$, $r_2$ and $r_3$. These keys and numbers should be kept unchanged in different task windows. In addition, the SP also need to choose different pseudonyms and a one-way hash function $H$.

\subsubsection{Task Request}
The SS first publishes some tasks including the requirements, the deadline and a possible range of credit $c_{i}\in[c_{min},c_{max}]$. If an SP decides to take part in the task $i$ ($i\in[1,M])$, he/she should communicate with the SS to request this task by using a pseudonym $PID_1$. Firstly, the SP negotiates the common information $i$ with the SS and requests RSA signature on the hash value of $RID$ to prove his/her identity in the credit deposit phase. Secondly, if the SS agrees with the common information, it returns the signature in an approval message:
\begin{equation}\label{1}
  SS \rightarrow SP: sign_{d}(H(RID))
\end{equation}
Then, the SP selects a random number $r_1$ and computes a request token identifier $ \tau_i=H(i||H(r_1))$ to ask for a partially blind signature on $\langle i,\tau_i\rangle$. The SS signs blinded $\langle i,\tau_i\rangle$ with its private key $K_1$ and returns the signature to the SP. So the SP can obtain $PBS_{K_1} (i,\tau_i)$ as the commitment and set $\gamma_i=\langle i,\tau_i,PBS_{K_1} (i,\tau_i)\rangle$ as the request token. Note that the SP can only obtain one request token with $\tau_i$ for the task. Finally, the SP sends the request token to the SS:
\begin{equation}\label{2}
   SP \rightarrow SS: PID_1, i,\tau_i, PBS_{K_1}(i,\tau_i)
\end{equation}
On receiving the SP's request token, the SS verifies the correctness of the signature $PBS_{K_1} (i,\tau_i)$. If it is valid, the SS sends the initial price $c_{max}$ to the SP. At the same time, the SS stores the used request token into its list to prevent the SP from reusing it.
\subsubsection{Report Submission}
In order to obtain the report token, the SP does as follows:

\begin{itemize}
     \item [-]Choose a random number $r_2$ and compute a credit token identifier $m_{ij}=H(i||j||H(r_2))||sign_d(H(RID))$, where $j=1,2,...,c_{max}$;
     \item [-]Select a random number $r_3$ and compute blind factor $z_{ij}=H(i||j||H(r_3))$;	
     \item [-]Generate the blinded message $\mu_{ij}=(m_{ij}\times z_{ij}^ e)$ $mod$ $q$. The report token identifier is $b_{ic}=H(\mu_{i1} ||\mu_{i2} ||...||\mu_{ic_{max}} ||i||c_{max})$;
     \item [-]Request a partially blind signature on $\langle i,b_{ic}\rangle$ using another random pseudonym $PID_2$.
\end{itemize}

Upon receiving the request from the SP, the SS signs blinded $\langle i,b_{ic}\rangle$ with its private key $K_2$ and delivers the signature to the SP. After removing the blind factor, the SP obtains $PBS_{K_2} (i,b_{ic})$ as the commitment. Therefore, the report token for this task is $\delta_{ic}=\langle i,b_{ic},PBS_{K_2} (i,b_{ic})\rangle$. Note that the SP can only get one report token with $b_{ic}$ for the task. Then, the SP encrypts the data report with the SS's public key $e$. The SP submits the report for task $i$, the report token, the blinded message and a timestamp:
\begin{equation}
\begin{split}
  SP \rightarrow SS:  &  PID_2,i,b_{ic},PBS_{K_2} (i,b_{ic}) \\
    & u_{i1},u_{i2},...,u_{ic_{max}},T_{i},E_e(report)
\end{split}
\end{equation}
Next, the SS verifies if $b_{ic}=H(\mu_{i1} ||\mu_{i2} ||...||\mu_{ic_{max}} ||i||c_{max})$ holds. If it dose, the SS knows that $\mu_{ij}$ $(j=1,2,...,c_{max})$ has also been committed for task $i$ and then verifies signature $PBS_{K_2}(i,b_{ic})$. If the signature is valid, it means that $b_{ic}$ has been committed for task $i$. Therefore, the SS stores the legal report token into its list. Furthermore, the SS decides to deliver a certain amount of credit $c$ $(c\leq c_{max})$ to the SP according to the task difficulty, data quality and other relevant factors. The SS signs $(\mu_{ij}\parallel T_{i})$ $(j=1,2,...,c)$ with its private key:
\begin{equation}\label{4}
   SS \rightarrow SP: sign_d(\mu_{i1}\parallel T_{i}),...,sign_d(\mu_{ic}\parallel T_{i})
\end{equation}
Upon receiving $sign_d(\mu_{ij}\parallel T_{i})$, the SP removes the blinding factor $z_{ij}^e$ $mod$ $q$ to get blind RSA signature $sign_d(m_{ij}\parallel T_{i})$. In this way, the SP obtains $c$ credit tokens $\varepsilon_{ij}=\langle m_{ij},sign_d (m_{ij}\parallel T_{i} )\rangle$, $(j=1,2,...,c)$. Without submitting the report, the SP cannot obtain any credit token.

\subsubsection{Credit Deposit}
\begin{figure}[tb]
\begin{center}
\includegraphics[width=8.5cm]{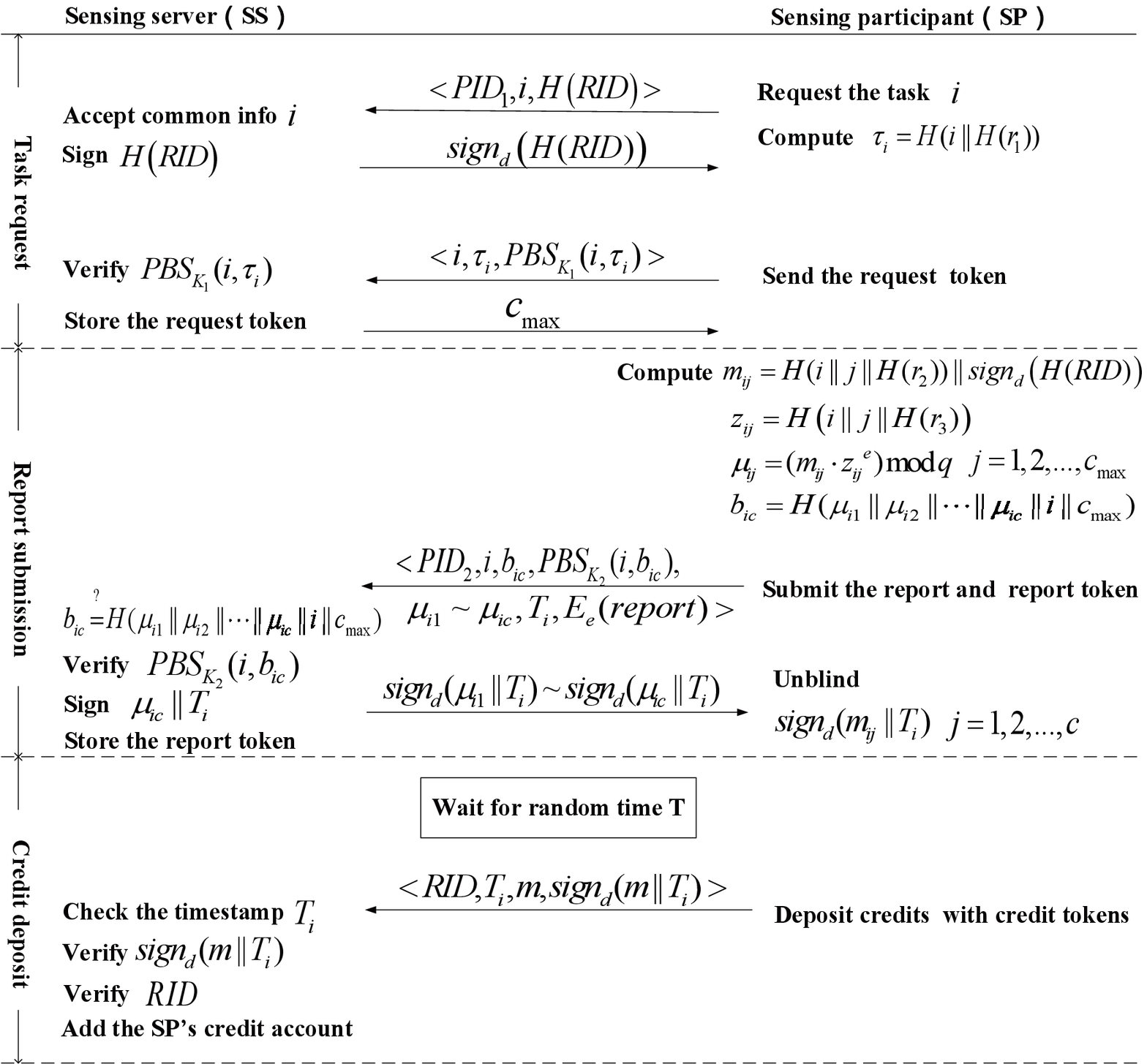}\\
\caption{Program flow chart of the proposed scheme}\label{scheme}
\end{center}
\end{figure}

After getting these credit tokens $\langle m_{ij},sign_d (m_{ij}\parallel T_{i})\rangle$, the SP can only submit one credit token at a time using his/her real identity $RID$. So, to mitigate timing attacks, all credit tokens need to be uploaded $c$ times in a random interval:
\begin{equation}\label{5}
  SP \rightarrow SS: RID,T_i,m,sign_d (m\parallel T_{i})
\end{equation}
After receiving the credit token, the SS verifies if the timestamp $T_i$ is expired. If it is not, the SS does the following steps: 1) verify the signature $sign_d (m\parallel T_{i})$ with its public key $e$; 2) verify $RID$ by extracting $sign_d(H(RID))$ from $m$. If two signatures are both valid, the SP's credit account can be added by one. At the same time, the SS stores these credit tokens into its list to avoid replay attacks. The flow chart of the above phases is illustrated in Fig. \ref{scheme}.

\subsubsection{Token Renewal}
For each task, the SS maintains a list of the used tokens to avoid reusing. When a task is completed, the corresponding request tokens and report tokens in the list should be released. However, the credit tokens will be released until all tasks in the same window have been finished or expired. Then, the SPs can request the following $M$ tasks whose indexes are $kM+1,kM+2,...,(k+1)M$ $(k \geq 0)$ if they are willing.

\subsection{Security Analysis}
\begin{figure}[tb]
\begin{center}
\includegraphics[width=8.5cm]{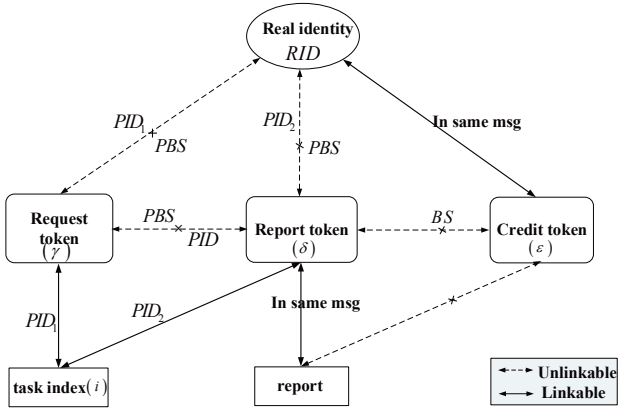}\\
\caption{The linkability between different components}\label{components}
\end{center}
\end{figure}

In this part, we analyze the security of the proposed scheme in terms of incentive and privacy. We first give the linkability between different tokens and objects in Fig. \ref{components}.
\subsubsection{Attacks on Incentive}

\

\noindent \textbf{Proposition 1.} \emph {The proposed scheme can prevent dishonest SPs from earning more than their due credits.}

\noindent \textbf{Proof:} {Since the SP binds the commitment in each request or report token to the task index through partially blind signature, he/she cannot use the request or report token for another task. Although the SP does not bind the credit tokens to the task index, the credit tokens can not be reused for another task in this window. Because the used credit tokens have been stored in the SS's list. Besides, the timestamp $T_{i}$ can resist replay attacks, so dishonest SPs can not reuse the credit tokens in the following task windows for more credits.}

\noindent \textbf{Proposition 2.} \emph {The proposed scheme can prevent malicious SPs from compromising the other SPs or forging the credit tokens to earn more credits.}

\noindent \textbf{Proof:} {A malicious SP may compromise the other SPs for their credit tokens, but he/she still cannot earn any credit. Since each credit token is associated with the SP's real identity $RID$, the SS will not offer the credit to the other SPs. Furthermore, it is impossible for the malicious SP to forge the signature $sign_d(H(RID))$, because the private key $d$ is kept secretly by the SS. Thus, the credit tokens $\langle m_{ij},sign_d (m_{ij}\parallel T_{i})\rangle$ cannot be forged. }

\subsubsection{Attacks on Privacy}

\

\noindent \textbf{Proposition 3.} \emph {The SS cannot link tasks to the corresponding SP.}

\noindent \textbf{Proof:} {Since the SP requests tasks with different random pseudonyms in the task request phase, so the SS cannot distinguish if two different pseudonyms belong to the same SP. Thus, the SS is unable to link the tasks to the corresponding SP.}

\noindent \textbf{Proposition 4.} \emph {The SS cannot link a report to the corresponding SP.}

\noindent \textbf{Proof:} {The SP uploads a report with a random pseudonym instead of his/her real identity in the report submission phase. Meanwhile, due to the untraceability of the partially blind signature, the SS cannot link the uploaded report to the corresponding SP's real identity, as shown in Fig. \ref{components}.

\noindent \textbf{Proposition 5.} \emph {The SS cannot link the credit tokens to a certain report.}

\noindent \textbf{Proof:} {Though, in the credit deposit phase, the SP submits the credit tokens with his/her real identity $RID$, the SS cannot link the credit token to the corresponding report token, because the connection between $m_{ij}$ and $u_{ij}$ is covered by $z_{ij}$ in blind RSA signature.

Within a short time, a task may be completed by one or a few SPs. In this case, if an SP deposits multiple credit tokens once, the SS may directly link the report to this SP. Thus, the SPs need to deposit one credit token at a time to prevent the SS from linking the credit tokens to the report.}

\section{Performance Evaluation}
In this section, we analyze the performance of the proposed scheme in terms of storage cast and computation overhead. For the experimental evaluation, the simulation environment is set up in Ubuntu 16.04 with an Intel(R) Core i5-4200U 1.60GHz $\times$2 processor and 2.2GB memory. The proposed scheme will be run 100 times in order to compensate for the randomness of the results.

\subsection{Storage Cost }
The SS needs to store a request token and a report token for one task until this task is finished. Also, it requires to store at most $M\times c_{max}$ credit tokens until all tasks in one task window are completed. Hence, the storage overhead is not heavy for the SS. After submitting the report for the task, the SP is required to store $c$ credit tokens delivered by the SS. By contrast, in \cite{li2014providing}, each SP needs storing $M(2\times c_{max}+1)$ commitments for the next $M$ tasks, so our scheme achieves lower storage cost for the SP.

\begin{table}[h]
  \centering
  \caption{The RUNNING TIME OF CRYPTOGRAPHIC PRIMITIVES} \label{primitives}
  \footnotesize
  \tabcolsep 0.05in
  \setlength{\extrarowheight}{0.10cm}
 \begin{tabular}{c|ccc}
    \hline
     \quad \raisebox{0.0cm}{} \quad \quad & \quad \raisebox{0.0cm}{PBS} \quad \quad & \quad \raisebox{0.0cm}{SIG} \quad  \quad & \quad \raisebox{0.0cm}{H} \quad    \\
    \hline
    SS  &   13.887ms &  5.037ms  &    3.421ms     \\   \hline
    SP  &    4.279ms &  0.469ms  &    0.015ms       \\

    \hline
 \end{tabular}
\end{table}

\begin{table}[h]
  \centering
  \caption{Time Consumption} \label{time}
  \footnotesize
  \tabcolsep 0.05in
  \setlength{\extrarowheight}{0.1cm}
 \begin{tabular}{c|c|c|c|c}
    \hline
    \raisebox{0.0cm}{} &  \raisebox{0.0cm}{Task application}  & \raisebox{0.0cm}{Report submission}  & \raisebox{0.0cm}{Credit deposit}  & \raisebox{0.0cm}{Total time}   \\
    \hline
    SS  &   17.924ms &  38.109ms &   4.705ms &   60.738ms\\\hline
    SP  &    4.793ms &  11.698ms &      -    &   16.698ms\\

    \hline
 \end{tabular}
\end{table}

\begin{figure}[tb]
\begin{center}
\includegraphics[height=7cm, width=8.5cm]{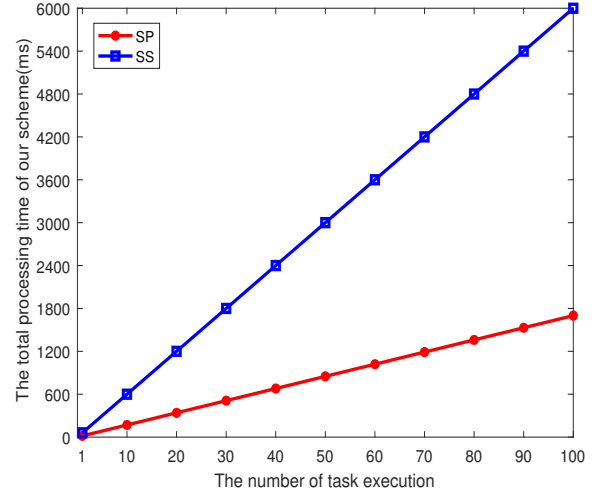}\\
\caption{Total processing time of our scheme vs. the number of task execution}\label{tasks}
\end{center}
\end{figure}

\subsection{Computation Overhead}
In the simulation procedure, the computation overhead is primarily caused by several major cryptographic operations. Table \ref{primitives} shows the running time of the cryptographic primitives. ``$PBS$" denotes a partially blind signature, ``$SIG$" denotes an RSA signature, and ``$H$" denotes a hash operation. Table \ref{time} indicates the processing time of each phase in our scheme when $M=1$ and $c=5$. On the SS side, the task request, report submission, and credit deposit phases only take several milliseconds. On the SP side, it just takes a short time for several one-way hash functions running in the terminal devices. Since the SP deposits one credit token each time in the credit deposit phase to protect participants' privacy, the SS has to sign more credit tokens, which may lead to a little more computation burden in the report submission phase. However, the processing time can be reduced if more efficient signature scheme is deployed instead of RSA signature. With the increase of the number of PSN tasks, the time consumption in the SS is still reasonable, as shown in Fig. \ref{tasks}. Additionally, according to the rough test, the communication overhead is about one thousandth of the computation overhead, so we ignore the communication overhead in the performance evaluation. In summary, our scheme takes reasonable computational overhead to meet the privacy-preserving inventive requirements in PSNs.

\section{Conclusion}
To facilitate the large-scale deployment of participatory sensing applications, we propose an efficient privacy-preserving incentive scheme without TTP to solve the conflict between participants' privacy and incentive in the participatory sensing networks. By combining of timestamp and nonce tokens, it can prevents replay attacks effectively, so dishonest users cannot reuse different tokens. Also, a malicious SS cannot link the reports or tasks to the corresponding SP. Security analysis indicates that the proposed scheme meets the security and privacy-preserving requirements in PSNs. Finally, the performance evaluation shows that the proposed scheme achieves lower computation and storage cost.
\section{Acknowledgment}
This work is supported by Natural Science Basic Research Plan in Shaanxi Province of China (No. 2016JM6057), the 111 Project (B08038) and Collaborative Innovation Center of Information Sensing and Understanding at Xidian University.

\ifCLASSOPTIONcaptionsoff
  \newpage
\fi

\bibliographystyle{IEEEtran}
\bibliography{ms}

\end{document}